\documentclass{jaa}
\usepackage{natbib}
\usepackage{graphicx}

\begin{document}\sloppy

%%paper title
%%For line breaks \\ can be used within title
%\title{Title of the paper goes here:\\ Second line}

\title{In orbit performance of UVIT over the 5 years}

%%author names are separated by comma (,)
%%use \and before the last author name
%%use a * along with the number separated by comma
%% for the  author for correspondence
%%\textsuperscript{number} is used for affiliation
%%\affilOne, \affilTwo etc., upto \affilTwentyfive is possible
%%Please note the first letter after \affil is capitalised in the command
%%

%\author{AUTHOR1\textsuperscript{1}, AUTHOR2\textsuperscript{1} and AUTHOR3\textsuperscript{2,*}}
%\affilOne{\textsuperscript{1}Department of P, University X, Place Pincode, Country.\\}
%\affilTwo{\textsuperscript{2}Department of Q, University Z, Place Pincode, Country.}

\author{S. K. Ghosh\textsuperscript{1,*}, P. Joseph\textsuperscript{2}, A. Kumar\textsuperscript{2},
 J. Postma\textsuperscript{3}, C. S. Stalin\textsuperscript{2}, A. Subramaniam\textsuperscript{2},
 S. N. Tandon\textsuperscript{4,2},
 I. V. Barve\textsuperscript{2},
 A. Devaraj\textsuperscript{2},
 K. George\textsuperscript{2,8},
 V. Girish\textsuperscript{5},
 J. B. Hutchings\textsuperscript{6},
 P. U. Kamath\textsuperscript{2},
 S. Kathiravan\textsuperscript{2},
 J. P. Lancelot\textsuperscript{2},
 D. Leahy\textsuperscript{3},
 P. K. Mahesh\textsuperscript{2},
 R. Mohan\textsuperscript{2},
 S. Nagabhushana\textsuperscript{2},
 A. K. Pati\textsuperscript{2},
 N. Kameswara Rao\textsuperscript{2},
 K. Sankarasubramanian\textsuperscript{7},
 P. Sreekumar\textsuperscript{5,2},
 \and S. Sriram\textsuperscript{2}}
\affilOne{\textsuperscript{1}Tata Institute of Fundamental Research, Mumbai 400005, India\\}
\affilTwo{\textsuperscript{2}Indian Institute of Astrophysics, Bangalore 560034, India\\}
\affilThree{\textsuperscript{3}University of Calgary, Calgary, Alberta, Canada\\}
\affilFour{\textsuperscript{4}Inter-University Centre for Astronomy \& Astrophysics, Pune 411007, India\\}
\affilFive{\textsuperscript{5}ISRO Headquarters, Bengaluru 560094, India\\}
\affilSix{\textsuperscript{6}National Research Council of Canada, Herzberg Astronomy and Astrophysics, Victoria, Canada\\}
\affilSeven{\textsuperscript{7}U.R. Rao Satellite Centre, Bengalure 560017, India\\}
\affilEight{\textsuperscript{8}Ludwig-Maximilians-Universit{\"a}t, Munich, Germany\\}
%\corres{swarna@tifr.res.in}
%%%%%%%%%%%%%%%%%%%%%%%%%%%%%%%%%%%%%%%%%%%%%%%%%%%%%%%%%%%%%%%%%%%%%%%%%%%%%%%%%%%%%%%%%%%%%%%%%%
%%escape two column mode for title, affiliation and abstract
%%by giving \twocolumn command as shown
\twocolumn[{
\maketitle
%%include \corres to print the corresponding author Email id
%\corres{abc@xyz.com}
\corres{swarna@tifr.res.in}
%%include \msinfo for
%%manuscript information such as
%%received, revised and accepted dates
%%
%\msinfo{1 January 2015}{1 January 2015}
%\msinfo{28 October 2020}

%%abstract
\begin{abstract}
%Abstract text goes here.  Abstract text goes here.
%Abstract text goes here.  Abstract text goes here.
% Abstract text goes here.
%Abstract text goes here.  Abstract text goes here.
% Abstract text goes here.
% Abstract text goes here.

Over the last 5 years, UVIT has completed observations of more than 500 proposals
 with $\sim$ 800 unique pointings.
%Observations for more than 500 proposals have been made by UVIT.
 In addition, regular planned
monitoring observations have been made and from their analysis various key parameters related to in
orbit performance of UVIT have been quantified. The sensitivities of the UV channels have remained
steady indicating no effect of potential molecular contamination confirming the adequacy of all the
protocols implemented for avoiding contamination. The quality of the PSF through the years confirms
adequacy of thermal control measures. The early calibrations obtained during the Performance
Verification (PV) phase have been further revised for more subtle effects. These include flat fields and
detector distortions with greater precision.
The operations of UVIT have also evolved through in orbit experience, e.g. tweaking of operational
sequencing, protocol for recovery from bright object detection (BOD) shutdowns, parameters for BOD
thresholds, etc. Finally, some effects of charged particle hits on electronics led to opimised strategy for
regular resetting. The Near-UV channel was lost in one of such operations. All the above in-orbit
experiences are presented here.

\end{abstract}

%%insert keywords separated by 3 hyphens using \keywords{words}
%\keywords{keyword1---keyword2---keyword3.}
\keywords{space vehicles: AstroSat -- telescopes: UVIT -- instrumentation: astronomical imaging}

}]
%%close the twocolumn escape here
\msinfo{}
%\msinfo{31 October 2020}

%%include \doinum{number}for the DOI number in the header
%%include \volnum{number} for the volume number in the header
%%include \year{yyyy} for  year of publication in the header
%%include \pgrange{num--num} page range of article in the header
%%include \artcitid{num} for the article citation id
%%include \lp to print last page of the article
%%include \setcounter{page}{pagenum} for the exact starting page of the article

%\doinum{12.3456/s78910-011-012-3}
%\artcitid{\#\#\#\#}
%\volnum{000}
%\year{0000}
%\pgrange{1--}
%\setcounter{page}{1}
%\lp{1}

\section{Introduction}

The Ultra-Violet Imaging Telescope (UVIT) is one of the five major scientific payloads on board the
first Indian multi-wavelength astronomical satellite mission AstroSat, which was launched on
September 28, 2015, with the Indian Space Research Organisation, ISRO’s PSLV-C30 rocket.
UVIT consists of two identical telescopes of aperture 375 mm and field of view $\sim$ 28${}^{\prime}$.
%%%
UVIT has high angular resolution imaging capability in the Far-UV (130 -- 180 nm) \& Near-UV (200 --
300 nm) wavebands using selectable narrow / medium / wide bandwidth filters as well as slit-less
spectroscopic imaging. A simultaneously viewing optical band, VIS (320 -- 550 nm) is incorporated
to aid implementation of the shift and add algorithm, for getting long exposure images from short
exposure frames, on the ground for avoidance of blurring due to drift in the telescope aspect.
The details about various sub-systems of UVIT and their respective qualifying tests and
 calibrations are described in Kumar et al. ~2012a \& Kumar et al. ~2012b.
%%%
 After an
initial 6 week long in orbit out-gassing phase, gradually individual subsystems of UVIT have been
tested for their functionalities and performance with the doors of the twin telescopes still closed. The
first light of the entire end to end UVIT system was carried out on November 30, 2015, by imaging the
Galactic open cluster NGC 188. This was followed by the nearly six month long Performance
Verification (PV) phase when detailed tests, characterization and calibrations were carried out. Many
details about results from the early phase of UVIT in orbit have been presented in Subramaniam et al.
(2016b), Tandon et al. (2017a) \& Tandon et al. (2017b). UVIT was thrown open (along with other
payloads for X-ray astronomy) for astronomical observations planned as per peer reviewed scientific
proposals, at first under
guaranteed time (GT) cycle followed by announcement of opportunity (AO) cycles.
During these Cycles, additional Calibration proposals have also been executed at regular intervals for
monitoring the health and quantifying the stability of UVIT’s performance. This article summarizes the journey of UVIT over the first five years in orbit in terms of performance and achievements including
unanticipated events and recovery therefrom.
%%%%%%%%%

\section{Performance Verification Phase}
%%%%
\subsection{Tests prior to opening of UVIT doors}

The earliest in orbit activities of UVIT pertained to qualification of all functions of the payload other
than the optical systems prior to opening of the doors, lasting $\sim$ 50 days post launch. (In these 50 days
only 17 days were used for operations with UVIT, rest were used by the other 4 instruments.) These
involved electrical and mechanical sub-systems -- e.g. communication systems with the spacecraft,
various ``states" of the Detector module, detector read out system, detector safety logic (Bright Object
Detection, BOD), generation of high voltages, rotational movements of the filter wheel system etc.
Tests were carefully planned in phases with gradually increasing complexity and with live control
ensuring options of aborting in case of encountering any abnormality (required scheduling \&
coordinating operations only during visibility of the spacecraft above main ground station). Cosmic
Ray Shower events passing through the detector system provided opportunity for checking certain
functionality in the absence of UV photons from the sky. Beginning with operations for one band
(among FUV/ NUV/ VIS) at a time, eventually simultaneous all 3 band operations were qualified. No
abnormality was encountered through this phase of commissioning of UVIT's detector and filter wheel
systems prior to opening of the doors.

\subsection{First light and lessons from early imaging operations}

The very first imaging of the sky with UVIT was carried out on 30 November 2015 targeting the
Galactic open star cluster NGC 188, whose coordinates (high Declination) offered the advantage of
optimal visibility through any season allowing long term follow up monitoring. Accordingly, selected
stars in this cluster were used as secondary standards for photometric calibration. 

%% R1 below ..................
%One of the most
%critical components of UVIT, viz., the Detector module, had several selectable parameters designed to
%provide adequate safety against damage due to unacceptably bright object in field.
% The detector for
%each band is configured as image intensifier which consists of a photo-cathode deposited on the
%window where primary electrons are generated by photo-electric effect by incident photons.
%%%%%%% New text below :
%{\bf
One of the most
critical components of UVIT is the Detector module employing a complex
 image intensifier system, operable in photon counting mode.
Key requirements driving this choice were : (1) to keep drift in pointing
 to $<<$1$^{{\prime}{\prime}}$ within an individual frame,
 the speed of reading out was required to be $>$10 frames/s, (2) given that the total
 number of UV photons detected in 0.1 s could be as low as 10, a read out noise of $<$1 (rms)
  was required. Therefore, read noise of CCD would not be acceptable. Further, a red-block filter
   with zero red leak would be required with a CCD, to completely block the longer wavelengths.
%}
%%%%%%%%%%%%%%%%%%%%%
The image intensifier configured for each band of UVIT consists of a photo-cathode deposited on the
window where primary electrons are generated by photo-electric effect by incident photons.
%%%%%%%%%%%%%%%%%
 These are
then multiplied by a large factor (gain) using a Micro-Channel Plate assembly, MCP, biased to
selectable high voltages. The stream of secondary electrons exiting the MCP are made to strike a
phosphor acting as anode generating optical light pulses. These pulses are detected by a CMOS imager,
Star250, with 512x512 pixels coupled through a fibre-optic taper. This imager is continuously read out
as individual frames during imaging operation. The MCP needs carefully planned protection against
exposure from bright objects since it can deliver only a limited amount of total charge in its operational
life. The areas of MCP having experienced prolonged exposure to high fluxes loses its electron
multiplying functionality leading to ineffective areas or even complete damage. In addition, the
excessive load on the high voltage supplies could damage them too. Accordingly, two key safety
features have been introduced - (i) a Bright Object Detect (BOD) logic has been implemented in the
onboard signal processing scheme which triggers a safety shutdown of the affected band and also raises
an alarm to the spacecraft for similar action in other bands; \& (ii) the initial imaging of any fresh sky
field or filter mandatorily begins with a very low gain gradually achieving the optimal setting (by
ramping up the high voltages at a selectable rate) which allows the BOD logic sufficient time to process
the incoming raw frames read out from the imager. The selectable parameters related to the triggering
threshold for the BOD logic are : pixel threshold, 1-D size along faster read-out axis of the frame, and the number of consecutive frames in which the pixel threshold has been exceeded over at least one
string of pixels of 1-D size. Given the long period of detector’s operations on ground during extended
tests and calibrations, significant experience existed regarding choices for these parameters.

%%%%
%%%%%%%% 13-Dec-2020 Referee#2 based changes in the para below ...

 In spite of
these knowledge base, an accidental BOD operation awaited the UVIT team on first light of VIS
channel ! 
%% new below
%{\bf 
It resulted from an erroneous choice of settings for the High Voltages - 
the set corresponding to Photon Counting mode was configured while attempting
to image in Integration mode. This oversight was corrected swiftly.
Very soon another BOD trigger occured due to a star in a field thought to the safe,
indicating error in estimation of `signal per photon' from the ground calibration.
The observed signal in orbit was higher, needing adjustment of the trigger threshold. 
%The details of implementation of the threshold 
%%Available dynamic range for the threshold
% necessiated exercising faster frame read out with stacking to retain the
%optimal effective integration time as planned originally.
%}
% 
%This was understood swiftly followed by revision of the strategy for its frame read-out (which
%included faster read-out to avoid BOD \& with image stacking to achieve the necessary integration
%time).
%%%%%%%%%%%
 Tweaking of all detector related parameters to optimal values were completed within days of the
first light. The selection of the band whose clock would be used as ideal Master Clock, i.e. in effect the
clock for all the three bands, as well as a consistence sequence of their configuration for imaging were
achieved shortly. This sequence being rather critical, its reliable implemention was achieved by
embedding these details in well designed Macros for operations. The initial ramping up of high
voltages was enforced for every change of filter.

 Another important realization during this phase was
the need for safeguarding the detector against bright stars while imaging in Window mode, i.e.
recording data only for part of the full field (selected size smaller than the full size of 512x512). This
risk was mitigated by introducing a mandatory full window mode imaging preceding the imaging with
a smaller window so as to ensure that BOD is triggered in case a bright object is present in the field
though outside the selected window. All such details were implemented in macros for imaging
operations. Eventually, this protocol was extended for even full window imaging, to bring in uniformity
in operations for simplicity.

%%%%%%%%%%%% Fig-1
%%%%\begin{figure*}
%%%\begin{figure} [th]
%%%%\begin{figure*} [th]
%%%\centering
%%%\includegraphics[scale=0.35]{Fig1.eps}
%%%%\includegraphics[width=1.9\columnwidth]{Fig1.eps}
%%%%\includegraphics[width=1.9\columnwidth]{Fig1d_VIS_BOD_nc.eps}
%%%\caption{An example of Bright Object Detect (BOD) trigger due
%%%to appearance of an unanticipated non-celestial object in UVIT’s
%%%field of view. One image frame in VIS band is displayed. The
%%%bright streak (near the bottom left corner) caused this BOD trigger.
%%%}

%%%\end{figure}
%%%%\end{figure*}
%%%%%%%%%%%%%%%%%%%%

\section{Experiences from long term operations of UVIT}

UVIT has been in regular use after its commissioning serving scientific as well as calibration
proposals. The operations involved imaging as well as slitless spectroscopy. Over the last $\sim$ 5 years,
more than 500 proposals have used UVIT. While most of the time UVIT has performed extremely well
%generally
 meeting planned design specifications, at times there have been occasional technical issues,
needing urgent review followed by action. The most frequently occurring event has been the automatic
shutdown of UVIT due to trigger of Bright Object Detect (BOD) logic. When any band of UVIT
encounters a brighter than programmed safe limit, it autonomously parks itself in a safe mode and
raises an alarm to the spacecraft bus, which in turn shuts down all the three bands of UVIT following a
safe sequence of operations. From the record of which band triggered the BOD, on ground the cause
for exposure to such a field is investigated. It may be noted that extreme care is taken while technically
approving any sky field to be observed with UVIT, which involves consideration of all cataloged bright
optical and UV objects in the target field. In most cases, either a human error or a large offset in
pointing have been identified. On a few rare occasions, brightening of a (variable) star of brightness
close to the safety limit or passage of a bright non celestial object (shining satellite) is responsible for
triggering the BOD. An example of the latter is displayed in Fig. 1. After each instance of BOD
triggered shutdown of UVIT, a well defined recovery protocol is followed to normalize the UVIT
bands after which regular operations can proceed. Typically about 3 - 4 incidents of BOD trigger were
encountered annually. It is noteworthy that the over-current in High Voltage units (described later) was
never triggered. This possibly points to the quality /ruggedness of the intensifier and the high voltage
generating circuits.

%%%%%%%%% Fig-1
%\begin{figure*}
\begin{figure} [th]
%\begin{figure*} [th]
\centering
\includegraphics[scale=0.40]{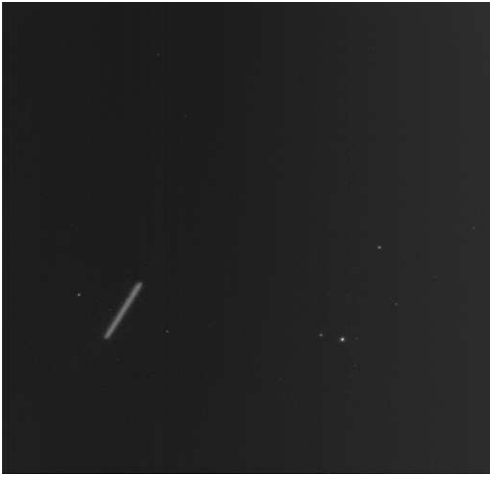}
\caption{An example of Bright Object Detect (BOD) trigger due
to appearance of an unanticipated non-celestial object in UVIT’s
field of view. One image frame in VIS band is displayed. The
bright streak (near the bottom left corner) caused this BOD trigger.
}

\end{figure}
%\end{figure*}
%%%%%%%%%%%%%%%%%

%%%%%%%%% Fig-2
%\begin{figure*}
\begin{figure} [th]
%\begin{figure*} [th]
\centering
\includegraphics[scale=0.40]{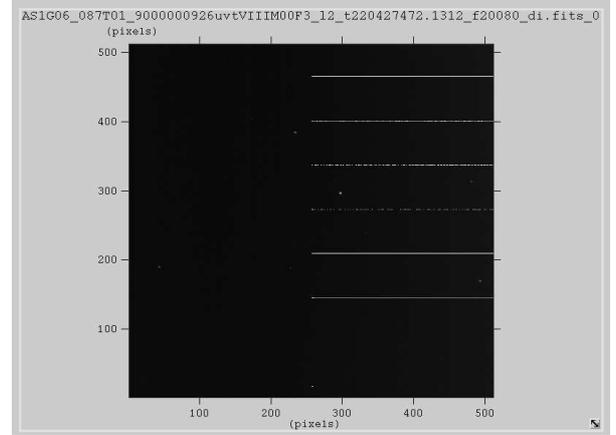}
\caption{
Example of artifacts appearing in VIS band images. The
horizontal stripes occur possibly due to effects of charged
particle hits and they disappear after a power reset.
}

\end{figure}
%\end{figure*}
%%%%%%%%%%%%%%%%%

An anomaly noticed rather early on (within months of in orbit operations) was appearance of stripes
in the raw image frames of VIS band (see Fig. 2). These were diagnosed to be effects due to charged
particle hits and complete recovery could be achieved by powering OFF the VIS electronics. Since the
processing pipeline on ground was agile enough to ignore such artifacts and generate final products
unaffected, the mitigation action was not carried out till the situation demanded powering OFF. At a
later phase, when such stripes in VIS appeared more often, along with some additional artifacts in UV bands too, a monthly schedule was drawn up to normalize (power OFF) all the 3 bands FUV, NUV \&
VIS.

%%%%%%%% 13-Dec-2020 : Para below modified based on Referee#2's comments 

The on board logic was designed to handle five kinds of anticipated emergency situations in UVIT.
%%%%%%
%{\bf Such conditions are defined based on : threats to the safety of intensifier of the detector \&
%high voltage power supplies;  anomalous status of the mechanism for the filter wheel; and
%operational error captured from invalid sequence of commands received. 
%The explicit list in increasing levels of severity is as follows :
%%They are (in increasing levels of severity) :
%}
%%%%%%%%%%%%%%%%%%%%%%%%%%%%%%%%%%%%
%{\bf 
% They are (in increasing levels of severity) :
 These are described in order of increasing level of severity.
 (1) TM Acknowledgment Error :
  Initiating an imaging session starting from inactive state,
involves configuring the Detector following an unique sequence of
commands which gradually activate relevant sub-systems ensuring complete safety. 
Another sequence (reverse order) is used for returning to inactive state.
Receipt of any command violating the above triggers this alarm.
%configure the Detector for imaging beginning from the normal
% inactive state, involves
%an unique sequence of commands which gradually activate sub-systems
%ensuring complete safety. Similarly for   
% a multi-step  change of  sequence of `State'-s  
% due to receipt of an
%‘illegal’ command - attempting a change of State for the Detector, which is disallowed;
% (2) Wrong ‘acquire’ status for the filter wheel – failed to reach targeted angle;
(2) Failure of the filter wheel mechanism to reach its targeted angle within a stipulated time.
 (3) Threat to safety of the intensifier on detection of a bright star (BOD).
%Detection Bright Object Detect trigger;
 (4) Threat to safety of the high voltage power supplies when current drawn exceeds the set limit.
% of High Voltage supply;
% and (5) Single Event Latchup event.
(5) Transition to Fail Safe state on detection of high current indicating
 radiation induced Single Event Latchup. 
%}
%%%%%%%%%%% 
%%%%%%%%
 One critical point for
each of these types of emergencies is the final parking state. During pre-launch deliberations, it was
decided that the relatively benign ones (first 3) should land the detector electronics (for all 3 bands) to
‘Low Power’ state and the remaining 2 severe ones to complete power ‘OFF’ state through operation of
mechanical relays. Being cautious during early in orbit operations, this strategy was made more
conservative by demanding all 5 situations to lead to OFF state. As a result, BOD used to result in OFF
state needing re-boot of the system during recovery operation.

During one of the post-BOD recovery operations (January 20, 2018), the NUV channel failed to
restart. Based on ground simulations using the Engineering Model and after prolonged trials based on
different well considered strategies (stretched RESET pulses), the NUV band could be recovered
(February 16, 2018). Unfortunately, failure of the NUV channel to boot recurred after a routine
monthly normalization (March 20, 2018). A very long struggle to revive the NUV channel followed.
The various innovative strategies employed to recover the lost channel included : (i) different
frequency of normalization trial, (ii) widening of the RESET pulse, (iii) gradual warming up of the
affected electronics and trying RESET at selected higher temperatures (this involved tweaking
parameters of the spacecraft's thermal system settings), etc. While none of the above could recover the
NUV channel, the main cause of the failure was eventually understood through deeper study of
technical literature.

%%%%%%%%% Fig-3
%\begin{figure*}
\begin{figure} [th]
%\begin{figure*} [th]
\centering
\includegraphics[scale=0.35]{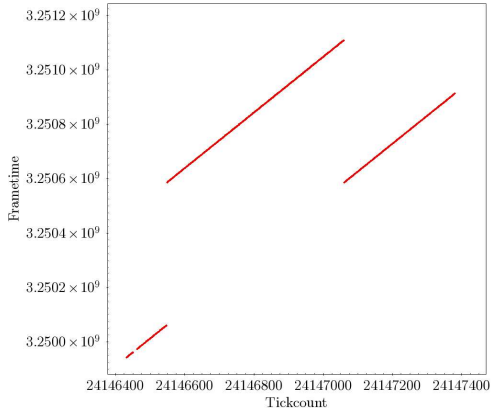}
\caption{
Example of an artifact due to a stuck bit in the Master
Clock. The time stamped on individual frames (‘Frametime’),
are plotted against the elapsed time (‘Tickcount’). The fixed
amount of jumps at each discontinuity indicates the 20 th bit of
the ‘Frametime’ to be stuck at zero.
}

\end{figure}
%\end{figure*}
%%%%%%%%%%%%%%%%%

Technical understanding of this failure emerged as follows : the onboard processing of the UVIT
detector data is implemented in a FPGA, whose code gets loaded from an EEPROM every time the
system is powered on or undergoes RESET. The EEPROM (\& also the FPGA) used were not of the
radiation hardened grade but of MIL spec grade, which is susceptible to damage by Cosmic Ray
radiation. In addition, the EEPROM was of serial type which is known to compromise the reliability.
As a further weakness, the code design did not provide any option for re-programming. Such devices
develop weak-cells (due to radiation damage) which are also known to deteriorate further with more
read-cycles.

Based on the above technical understanding of the reason for loss of NUV band, the practice of
monthly normalization of VIS \& FUV bands was completely discontinued. In addition, the strategy for
on board handling logic post occurrence of BOD emergency state was revised. The parking state on
BOD trigger was changed to ‘Low Power’ (avoiding any booting action which involves re-loading of
the code from EEPROM to FPGA). Attempts for reviving the NUV band continues to date at regular
intervals.

%%%%%%%%% Fig-4
%\begin{figure*}
\begin{figure} [th]
%\begin{figure*} [th]
\centering
\includegraphics[scale=0.35]{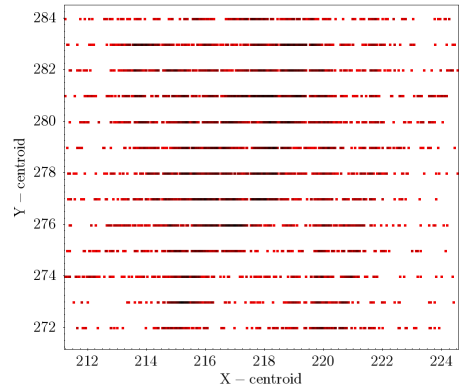}
\caption{
Example of multiple bits of the centroid coordinates (along
Y-axis) of every photon event to be stuck to zero. The points in
the plot correspond to individual photons. The systematic gaps
in the values of Y-centroid indicate several successive least
significant bits (corresponding to the fractional part) are affected.
}

\end{figure}
%\end{figure*}
%%%%%%%%%%%%%%%%%

More recently (since $\sim$ December 2018), some other types of artifacts due to radiation hits were
discovered, which could be recovered. However, their recovery required power RESET. These
artifacts were : (a) frozen 20th bit of VIS clock which is the selected Master for all bands (see Fig. 3),
(b) certain frozen bits of all X-centroids for photon events, etc. Initially, on encountering (a), plans and
procedures were set up for change over of the Master Clock from VIS band to FUV band – thereby
avoiding issuance of any RESET (in view of the loss of NUV band). This involved re-programming all
 command tables corresponding to `Imaging Parameters' for each band. The uploading of these tables to
the relevant spacecraft sub-system involved certain new activities. After a review by ISRO experts, this
course of action was not found to be advisable. In the mean time, (b) was encountered which forced the
use of RESETs which eventually mitigated both these issues. Most recently (October 2020), all bits
corresponding to fractional part of the Y-centroid were stuck (see Fig. 4). All instances of appearance of
such artifacts (other than periodic appearance of stripes in VIS image frames) and subsequent recovery
from them are presented chronologically in Table 1.

%%%%%%%%%%%%%%%%%% TABLE here (below) !!!!!!!!!!!!!!!!!!!!!!!!!!!!!!!!
%%Use table* environment to get the table spanning both the columns

\begin{table*}[htb]
\tabularfont
\caption{Chronology of issues observed in UVIT data* and their resolution}\label{firstTable}
\begin{tabular}{|l|l|l|l|l|}
\topline
%\textbf{head1}&\multicolumn{11}{c}{\textbf{head2}}&\textbf{head3}\\
\textbf{Serial}&\textbf{Date of}&\textbf{Date of}&\textbf{Type of issue}&\textbf{Remarks}\\
\textbf{ No.}&\textbf{Appearance}& \textbf{Resolution}&&\\
%\midline
\hline
%one& two &three&four&five&six&seven&eight&nine&ten&eleven&twelve&thirteen\\
%1&2&3&4&5&6&7&8&9&10&11&12&13\\
%aaa&bbbb&cccc&ddddd&eee&ffff&ggggg&hhhhhhhh&iiii&kkkkkk&hhh&jjjjjj&lllll\\
1 & 16-Apr-2017 & 06-Jun-2017 & Vertical stripes in& Recovered by RESET \\
  &             &             & NUV images         &  \\
\hline
2 & 29-Apr-2017 & 06-Jun-2017 & Anomalous size of & Recovered by RESET \\
%{\bf 2} & {\bf 29-Apr-2017} & {\bf 06-Jun-2017} & {\bf Anomalous size of}  & {\bf Recovered by RESET} \\
  &             &             & VIS images    & Data made usable by  \\
%  &             &             & {\bf VIS images}    & {\bf Data made usable by}  \\
  &             &             &          & developing mitigation  \\
%  &             &             &          & {\bf developing mitigation}  \\
  &             &             &          & scheme in ground software  \\
%  &             &             &          & {\bf scheme in ground software}  \\
\hline
3& 18-Nov-2017& 23-Nov-2017& No FUV data& Recovered by RESET \\
\hline
4& 06-Feb-2018& 22-Feb-2018 & Unable to turn on & Recovery managed\\
 &            &             & NUV band after a & after exploration\\
 &            &             & shutdown due to & of many strategies\\
 &            &             & BOD             & to power on \\
\hline
5& 30-Mar-2018& ---& Unable to turn on & No recovery yet,\\
 &            &    & NUV band after & despite repeated\\
 &            &    & a shutdown     & attempts exploring \\
 &            &    &                & multiple strategies\\
 &            &    &  & (periodic attempts continue)\\
\hline
6& 25-Dec-2018& 13-Jul-2019& Stuck 20th bit in & Eventual recovery\\
 &            &            & time stamps & by RESET (after \\
 &            &            & on FUV frames & adopting many \\
 &            &            &               & alternate schemes \\
 &            &            &               & avoiding RESET, \\
 &            &            &           & in view of the loss \\
 &            &            &           & of the NUV band); \\
 &&&& The data collected during \\
 &&&& this period made usable \\
 &&&& by developing mitigation \\
 &&&& scheme in ground software \\
\hline
7 & 05-Mar-2019 & 04-Jul-2019 & X-centroids (FUV) & Recovered by RESET \\
 &  &                         & with stuck bits & (after hesitation for \\
 &  &                         &                 & use of the RESET) \\
\hline
8 & 03-Nov-2019 & 16-Nov-2019 & Stuck 31st bit of & Recovered by RESET \\
 & &                         & Time Stamp in FUV & (after exploring  \\
 & &                         & band frames (relaying & alternatives to RESET) \\
 & &                         & Master Clock from VIS) & \\
\hline
9 & 26-Nov-2019  & 13-Dec-2019 & Stuck 31st bit of & Recovered by RESET \\
  &  &                         & Time Stamp in FUV & (with understanding of \\
  &  &                         & band frames (relaying & the damage, future \\
  &  &                         & Master Clock from VIS) & occurrences of the \\
  &  &                         &                       & issue became predictable; \\
  &  & &                                         & devised a periodic RESET \\
  &  & &                                         & plan, which works well) \\
\hline
10 & 13-Oct-2020 & 15-Oct-2020  & Y-centroids (FUV)  & Recovered by RESET \\
 & &  &  with  stuck bits &  \\
\hline
\end{tabular}
\tablenotes{*The many instances of appearance of stripes in VIS band images are not listed here. They were always resolved through RESET. Also mitigating scheme was incorporated in the Level-2 pipeline to handle such artifacts with no loss of functionality, so that data preceding RESET also remain fully usable.}
\end{table*}
%%%%%%%%%%%%%%%%%%%% !!!!!!!!!!!!!!!!!!!!!!!!!!!!

The lone instance (over 5 years of operations) of UVIT bands remaining in active imaging mode
beyond the mission schedule and being exposed during one full bright (sunlit) part of the orbit, was
experienced on September 19, 2020. However, no damage or degradation of any performance of UVIT
was noticed after this incident. The reason for this anomaly was traced to a logic in spacecraft
operations which has been mitigated.

One non-recoverable artifact encountered (early November 2019) was the stuck 31st bit of FUV clock
relaying Master clock counter from VIS band (which continued to record correct values). It is fortunate
that this bit can change state only after a few weeks and accordingly, this defect could be by passed by
an aggressive mitigation plan. The plan involved a periodic RESET of the VIS band electronics every 
$\sim$ 12 days.

Despite the anomalies described above, the two bands of UVIT, viz., FUV \& VIS have been serving
scientific observation plans leading to interesting research results.

%%%%%%%%% Fig-5
%\begin{figure*}
\begin{figure} [th]
%\begin{figure*} [th]
\centering
\includegraphics[scale=0.35]{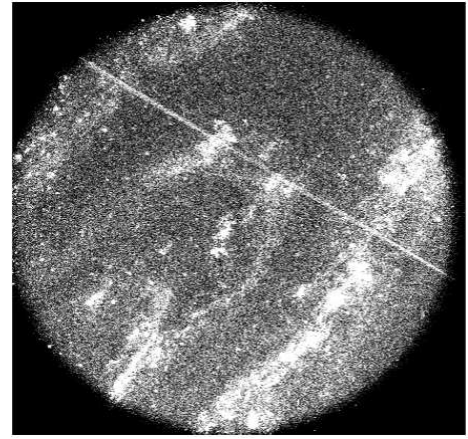}
\caption{
An unexplained feature (streak) observed in images in all
3 bands of UVIT for certain targets. As an example, image of the
galaxy M31 in NUV band is displayed. The orientations of the
streaks with reference to mountings of FUV / NUV / VIS detectors
imply that they could be caused by some structure within the
telescope tube.
}

\end{figure}
%\end{figure*}
%%%%%%%%%%%%%%%%%

Additional effects / events, some unexpected others expected, that were experienced are summarized
here.

The effects due to Cosmic Rays (CR) on the detector system were anticipated and their handling by
the offline data processing pipeline on ground were planned accordingly. The primary CR - energetic
charged particle - itself interacting in an imager pixel could corrupt its value corresponding to a large
signal. On the other hand, showers of secondary charged particles generated by interaction of primary
CR in the proximity of the detector and high electrical fields around MCP, mimic UV photons. While in
the low gain operation of the detector (employed for VIS band), only one pixel is affected per primary
CR, a large number of randomly located background events are recorded in the high gain operation of
the detector used for FUV \& NUV bands due to the showers. Fortunately, each shower lasts at most a
few micro-seconds, and hence can affect only one frame. However, the frequency of occurrence of such
cosmic ray shower is crucial. The mitigation plan in ground software is to identify frames affected by
showers using selectable parameters (statistically determined threshold on number of events) and
flagging them for discarding. From long term experience, on average 3.5 showers per sec are observed
contributing to $\sim$ 150 events /sec for full field operations. The impact of discarding affected frames is
rather small (loss of $\sim$ 10\% of the observations with full field). Given the uncorrelated nature of the
cosmic ray background events, they may be ignored for fields which are UV bright. However, for very
deep observations it is important to discard affected frames (Saha et al., 2020).

Example of an observed artifact is a bright ``streak" due to some very bright object within a few
degrees of the telescope axis (though outside UVIT’s 28 arc-min field of view). Such streaks have been
observed in VIS as well as NUV images. The orientation of the streaks in these images imply a
direction fixed to the telescope tube, once the relative angle between the axes of the detectors is taken
into account. Such streaks have been observed while observing the targets : M 31, PKS 1510-089 \&
Crab. One example is presented in Fig. 5 which is an NUV image of M 31.

%%%%%%%%% Fig-6
%\begin{figure*}
\begin{figure} [th]
%\begin{figure*} [th]
\centering

%%%
%\begin{subfigure}{\textwidth}

\includegraphics[scale=0.30]{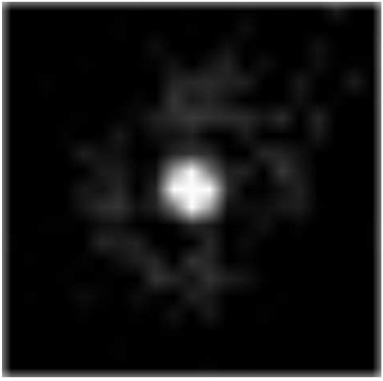}
%%\includegraphics[scale=0.35]{Fig6a.eps}
%\includegraphics[width=1.1\columnwidth]{Fig6a.eps}
%\includegraphics[width=1.1\columnwidth]{Fig6_NUV_sat_nc_p_l1.eps}

%\end{subfigure}

%\begin{subfigure}{\textwidth}

\includegraphics[scale=0.30]{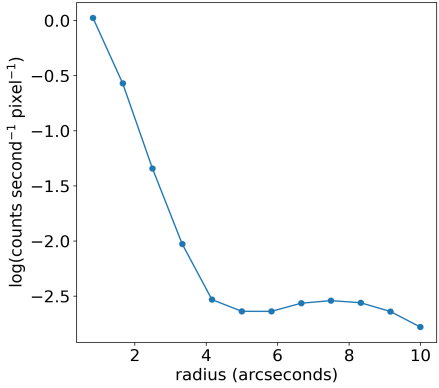}
%%\includegraphics[scale=0.35]{Fig6b.eps}
%\includegraphics[width=1.3\columnwidth]{Fig6b.eps}
%\includegraphics[width=1.3\columnwidth]{Fig6g_fin_sat_rad_prof_N188_nc.eps}
%\includegraphics[width=1.1\columnwidth]{Fig6g_N_sat_rad_prof_N188_nc.eps}

%\end{subfigure}

%%%%%%%%%%%%%%%
\caption{
Effect of saturation is illustrated in the NUV band image of
a bright star in the cluster NGC 188
%{\bf
 (size : 28$^{{\prime}{\prime}} \times$ 28$^{{\prime}{\prime}}$).
%}
 The valley encircling the
central peak is due to saturation effect of photon counting mode.
The radial profile displays the local dip quantitatively.
}

\end{figure}
%\end{figure*}
%%%%%%%%%%%%%%%%

Another artifact presented here relates to the effects due to saturation in photon counting mode of
imaging. A deep valley encircling the central peak is observed for a bright star in the star cluster NGC
188 imaged in NUV, which is displayed in Fig. 6. 
The accompanying plot of the radial profile quantifies this dip.

The ground segment software system at ISRO provides absolute time information for the science
data by correlating internal clocks of UVIT (one per band) with Universal Time Clock (UTC) in the
Level-1 (L1) products. This utilizes simultaneous samples of spacecraft’s clock counter with those from
UVIT. Accordingly, the UVIT’s Level-2 (L2) processing pipeline was designed using UTC as the
primary timing reference. However, often this time correlation in L1 was found to be unreliable. Hence,
it was necessary to incorporate new functionality in the L2 pipeline to by-pass UTC and use the Master
clock of UVIT which ensured inter-band time synchronization. Even in the UTC by-pass mode a
provision for approximate (good to $\sim$ 1 sec) absolute time (MJD\_UT) for every frame was made based on
an intelligent algorithm.

\section{Key Performance Parameters}

The key parameters of performance of UVIT include the following : (1) the photometric
calibration quantified by zero-point magnitude and the unit conversion factor for all the filters, (2)
effects of saturation, (3) variation of sensitivity across the field of view (flat field), (4) Point Spread
Function, PSF, and its variation over the field, (5) dispersion, resolution and effective areas in the
grating mode, and (6) astrometric calibration including distortion. The procedure followed to arrive at
these and related details have been presented in Tandon et al. (2017c) \& Tandon et al. (2020). Here we
summarize the results reported there.

Based on the observations carried out during the initial 18 months or so led to the first phase of
calibration results (Tandon et al. 2017c). 
The zero-point magnitudes for all the filters in FUV and NUV band were quantified.
% The zero-point magnitudes for the CaF2 filter (F148W) in the
% FUV band and the Silica filter (N242W) in the NUV band are 18.097 $\pm$ 0.01 \& 19.763 $\pm$ 0.002
%respectively. 
The measured sensitivities in FUV and NUV are found to be $>$ 80\% of the expectations
based on tests carried out on ground. The spatial resolutions (PSF FWHM $\sim$ 1.3 -- 1.5 arc-sec in FUV
and $\sim$ 1.0 -- 1.4 arc-sec in NUV) are found to be better than expected. The variation of PSF across the
28${}^{\prime}$ field is small for the FUV band. For the NUV band an increase of $\sim$ 10\% is found in FWHM in the
central part of the field compared to the edges. No detectable change in the PSFs in both UV bands
have been found to date. The astrometric accuracy over the full field is found to be $\sim$ 0.5 arc-sec RMS.

With passage of time ($\sim$ 3 years) additional regular calibration observations with UVIT were carried
out. Based on these extensive additional database and improved understanding of the instrument,
further refinements to the first phase calibrations were conducted.
The results from these studies have been reported in Tandon et al. 2020. 
 These improvements have led to
quantification of new photometric calibrations which included subtle effects.
% (Tandon et al. 2020).
The zero-point magnitudes, ZP, for most filters have been revised, and for some with improved precision
 (e.g. error on ZP for the filter N245M reduced from 0.07 to 0.005).
For example, improved determination of ZP for the CaF2 filter (F148W) in the
 FUV band and the Silica filter (N242W) in the NUV band led to the values 18.097 $\pm$ 0.01 \& 19.763 $\pm$ 0.002
respectively.

 The flat fields have been significantly
improved by supplementing remainders by analytic functions (third-order polynomial at the central
region \& linear in radius with azimuthal dependence for the outer regions).
%%%
The achieved accuracy with this improved flat field correction scheme has been estimated 
from exposures on multiple fields of Small Magellanic Cloud.
The fractional differences in the flat-field corrected counts for sources when
 they fall near centre of the
  field and when they fall near edge of the field are $\sim$ 0.06.
 This suggests that the errors on
flat-field corrections are no more than 6\%.
%A major contribution to this limit could be from Poisson statistics of the counts and errors of
%photometry near the edge.
%
%Based on exposures on multiple fields of Small Magellanic Cloud, the fractional errors on
%the final flat fields are $\sim$ 0.06,  a major part of which
%could be from Poisson statistics of the counts and errors of
%photometry near the edge.
%%%%%%%
% The distortions of the detectors have been characterized better. 

As a result of these improvements, the astrometric accuracy improved to 0.4 arc-sec (rms),
indicating uncorrected distortion to be $<$ 0.3 arc-sec (rms). 
%The flat fields have been significantly
%improved by supplementing remainders by analytic functions (third-order polynomial at the central
%region \& linear in radius with azimuthal dependence for the outer regions).
 The spectral, PSF and
astrometric calibrations have also been improved upon.
For both the NUV and FUV bands, the FWHM of the PSF is found to be 1.4 arc-sec or better
within the central 24 arc-min of the field.
 The new results conclude that there has been no reduction in sensitivities of FUV and NUV bands.

%%%%%%%%% Fig-7
%\begin{figure*}
\begin{figure} [th]
%\begin{figure*} [th]
\centering
\includegraphics[scale=0.35]{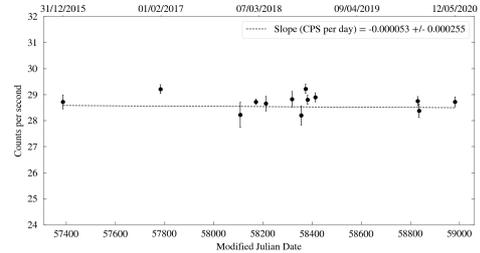}
\caption{
The sensitivity of the FUV band over its operations
covering earliest operations till May 2020 has remained
unchanged. The plot shows count rates from monitoring of a
secondary calibrator, WOCS-5885, – a star in the cluster NGC 188.
%{\bf
 The rms scatter is $\sim$ 1\% over 1600 days.}
%
%}

\end{figure}
%\end{figure*}
%%%%%%%%%%%%%%%%%

The result from monitoring the sensitivity of the FUV band over the entire mission including recent
observations is presented in Fig. 7, which shows no detectable change in the count rate for
a secondary calibrator star, WOCS-5885, in the cluster NGC 188 (Subramaniam et al., 2016a).
 The stability of sensitivities
over long duration has vindicated the care taken over the years on ground
 towards control of molecularcontamination, viz., choice of materials,
 operations in clean room, cleaning protocols, pre-assembly
baking, purging of the optical cavity of UVIT with pure N2 gas. In addition, the in orbit protocols
followed : long in orbit wait for degassing before opening the doors as well as avoiding direct sun light
falling on telescope tubes during spacecraft maneuvers also helped. The stability of the PSF is ensured
by adequate on board thermal control through the mission. The temperatures of critical elements, viz.,
telescope tubes and detectors (measured over $\sim$ a week) responsible for the observed stability in
sensitivity as well as PSF size are displayed in Fig. 8 \& Fig. 9 respectively. The temperature of the tubes
is stable within $\pm$ 0.5 C, which translates to a geometrical blurring in the image by $<$ 0.1 arc-sec rms,
implying a change from $\sim$ 0.5 arc-sec to $<$ 0.51 arc-sec rms for the PSF. The temperature of detector
(Star250 imager) is stable to within $\pm$ 1 C.

%%%%%%%%% Fig-8
%\begin{figure*}
\begin{figure} [th]
%\begin{figure*} [th]
\centering
\includegraphics[scale=0.35]{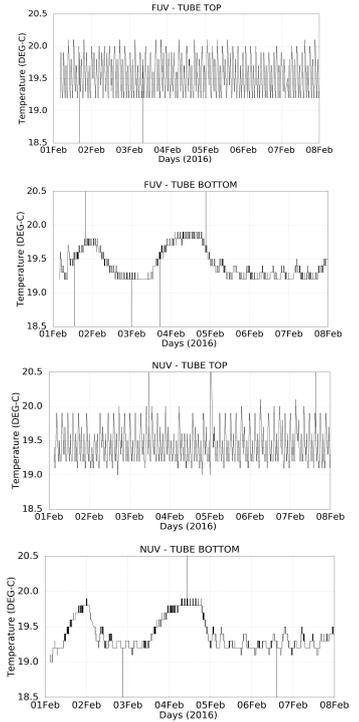}
\caption{
Example of typical variation of temperature of the two
UVIT Telescopes (FUV and NUV/VIS) at respective Tube Top
(TT) \& Tube Bottom (TB) over one week.
}

\end{figure}
%\end{figure*}
%%%%%%%%%%%%%%%%%

%%%%%%%%% Fig-9
%\begin{figure*}
\begin{figure} [th]
%\begin{figure*} [th]
\centering
\includegraphics[scale=0.30]{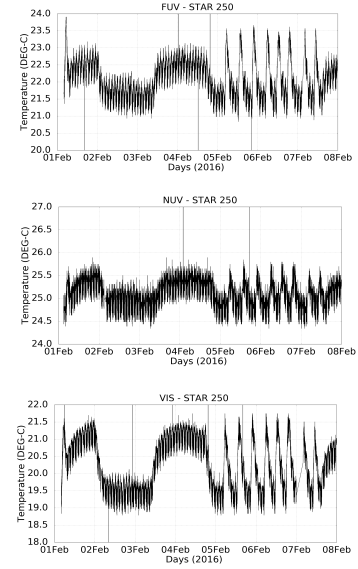}
\caption{
Example of typical variation of temperature of the
3 Detectors (FUV, NUV, VIS) in UVIT over one week.
%{\bf
The temperature is measured on the Camera Proximity Unit
housing the image Intensifier and the CMOS sensor STAR 250. 
%}
%
}

\end{figure}
%\end{figure*}
%%%%%%%%%%%%%%%%%

\section{Epilogue}

UVIT has been performing quite satisfactorily over the years with most its in orbit specifications
close to and a few even better than the corresponding targeted values. A large fraction of AstroSat time
has been allocated to UVIT by the Time Allocation Committee based on user driven scientific
proposals. The distribution on sky of the astronomical targets observed with UVIT is
 displayed in Fig. 10, which includes about 800 unique pointings.
 A large number of important astronomical results have
already appeared in journals with high impact (possibly many more are in the process).
% It may be noted
%that
 All scientific payloads of AstroSat including UVIT, had a baseline design life of 5 years, which has
already been achieved. While it is a pity that the NUV band was lost at the mid-point of this life, due to
radiation damage of some critical electronic components, the FUV \& VIS bands continue to serve the
users.
% It may be noted that 
%UVIT’s FUV band remains the only instrument in orbit currently providing
%unique opportunity for observations and serving the global astronomy community.
 It is hoped that these
two bands will last many years in future.
%{\bf
Although HST allows extremely sensitive imaging in the FUV (STIS, FUV-MAMA;
 25$^{{\prime}{\prime}} \times$ 25$^{{\prime}{\prime}}$),
 UVIT currently provides the unique
 opportunity for wide field imaging with 28$^{\prime}$ dia field.
%}

%%%%%%%%% Fig-10
%\begin{figure*}
\begin{figure} [th]
%\begin{figure*} [th]
\centering
\includegraphics[scale=0.45]{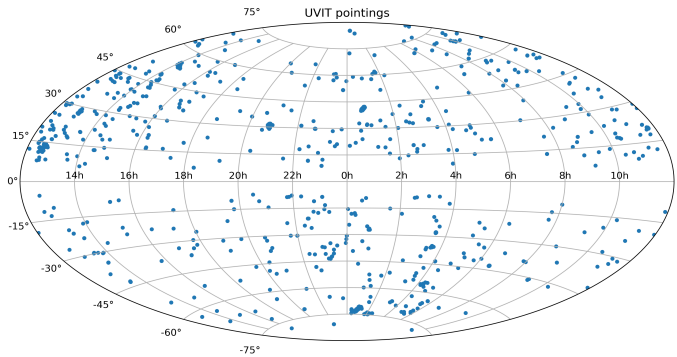}
\caption{
A plot showing distribution on sky of the astronomical
targets observed by the UVIT (total $\sim$ 800 unique pointings).
}

\end{figure}
%\end{figure*}
%%%%%%%%%%%%%%%%%

%\vspace{-2em}
%\section{Conclusion}
%Conclusion here.

%%Use section* for acknowledgements
\section*{Acknowledgements}

%Acknowledgements here.

The UVIT project is a result of collaboration between IIA, Bangalore, IUCAA, Pune, TIFR, Mumbai,
many centers of the Indian Space Research Organization (ISRO), and the Canadian Space Agency. We
thank these organizations for their support. We gratefully thank members of the Ground Segment
software and Mission Operations teams of ISRO for their continuing support. We also thank members
of the AstroSat Project and the AstroSat Science Working Group for their feedback.

\vspace{-1em}

%%use \balance somewhere in the left column of the last page to balance the two columns in the end page

%%References section

%
\end{document}